\documentclass[%
 reprint,
superscriptaddress,
 amsmath,amssymb,
 aps,
]{revtex4-2}

\usepackage{graphicx}
\usepackage{dcolumn}
\usepackage{bm}
\usepackage{hyperref}
\hypersetup{colorlinks,linkcolor={blue},citecolor={blue},urlcolor={black}}

\begin{document}

\preprint{APS/123-QED}

\title{Diversity of Ultrafast Spin Dynamics Near the Tricritical Point in a Ferrimagnetic Gd/FeCo Multilayer}
\author{T.G.H. Blank}
\affiliation{Radboud University, Institute for Molecules and Materials, 6525 AJ Nijmegen, the Netherlands}
\affiliation{Department of Applied Physics, Eindhoven University of Technology, P.O. Box 513, Eindhoven 5600 MB, the Netherlands}
\author{B.D. Muis}
\affiliation{Kavli Institute of Nanoscience, Delft University of Technology, 2600 GA Delft, the Netherlands}
\author{T. Lichtenberg}
\affiliation{Department of Applied Physics, Eindhoven University of Technology, P.O. Box 513, Eindhoven 5600 MB, the Netherlands}
\author{B. Koopmans}
\affiliation{Department of Applied Physics, Eindhoven University of Technology, P.O. Box 513, Eindhoven 5600 MB, the Netherlands}
\author{A.V. Kimel}
\affiliation{Radboud University, Institute for Molecules and Materials, 6525 AJ Nijmegen, the Netherlands}

\date{\today}
\begin{abstract}
It is found that subtle changes in the external magnetic field and temperature result in dramatic changes in the ultrafast response of spins to a femtosecond laser excitation in a ferrimagnetic Gd/FeCo multilayer. A total of six distinct types of spin dynamics were observed and explained by considering the spin-flop transition to the noncollinear phase and the concept of a tricritical point in the $H$-$T$ phase diagram. A particularly interesting type of dynamics is the exchange-driven reversal. These exchange-driven dynamics provide new insights into the tricritical point, which is shown to separate two thermodynamically distinct noncollinear phases with the transition-metal magnetization pointing on adjacent sides of the anisotropy plane. 
\end{abstract}

\maketitle

Ultrafast magnetism is a rapidly developing field of physics that explores spin dynamics launched in magnetically ordered materials by ultrashort stimuli with a duration of just a few ps or less. Such stimuli bring the magnetic media into a strongly non-equilibrium state where the conventional thermodynamic description of magnetic phenomena fails, often resulting in counter-intuitive spin dynamics~\cite{PhysRevLett.76.4250, Eschenlohr_2018, KIMEL20201}. Better fundamental understanding and exploring the practical limits on the timescales to control the magnetic ordering is expected to have a great impact on the development of magnetic information processing and data storage technologies~\cite{PhysRevLett.99.047601, Kimel2019}. 

Among all classes of magnetic materials studied in ultrafast magnetism (ferro-, ferri-, and antiferromagnets), ferrimagnets are most promising for the ultrafast and efficient control of magnetic properties. In these materials, the exchange interaction aligns the spins antiparallel as in antiferromagnets, but the spins are not equivalent such that the net magnetization is nonzero. Examples of ultrafast phenomena discovered in ferrimagnets include magnetization reversal via a strongly non-equilibrium state initiated by fs laser pulses~\cite{Radu2011, PhysRevApplied.13.024064}, current-induced magnetization reversal~\cite{doi:10.1126/sciadv.1603117}, as well as the record fast and least dissipative magnetic recording~\cite{Stupakiewicz2017}. A central ingredient that enriches ultrafast magnetism in ferrimagnets is the fact that the balance between the spins and therefore their magnetic structure changes with temperature. In particular, a \textit{compensation temperature} may exist where the spins cancel out and the net magnetization is zero. Ferrimagnets in the vicinity of their compensation temperature combine the properties of ferro- and antiferromagnets and therefore offer a highly intriguing playground in ultrafast magnetism~\cite{Mangin2014, Kim2022, hintermayr2023exploring}. 

Lesser known is that, next to temperature, the magnetic structure of ferrimagnets can also be tuned with an external magnetic field, resulting in a rich $H$-$T$ phase diagram~\cite{Zvezdin1998}. When a critical field is reached, the spins undergo a \textit{spin-flop transition} to a noncollinear spin configuration \cite{10.1063/1.1656100}. Previous studies of ferrimagnetic rare-earth (RE) transition-metal (TM) alloys have indeed shown that sufficiently high external magnetic fields unlock a new dimension in the ultrafast magnetism of ferrimagnets related to this noncollinear phase~\cite{PhysRevLett.118.117203, PhysRevB.100.174427}. However, due to relatively strong exchange interaction between the sublattices, the studies required exceptionally high magnetic fields up to $30$~T. Here, we overcome this obstacle by fabricating a high-quality synthetic ferrimagnet -- a Gd/FeCo multilayer. In such a heterostructure, the amount of nearest neighbors from different species (RE/TM) decreases as compared to the alloy, meaning that the RE-TM exchange interaction, as well as the critical field, are expected to drop significantly \cite{PhysRevB.104.054414}. In this Letter, we show in Gd/FeCo that by changing the temperature or magnetic field in relatively narrow ranges, from $320$~K to $340$~K and from $0.5$~T to $1$~T, we can reach all possible magnetic phases hosted by ferrimagnets and trigger largely diverse ultrafast spin dynamics upon femtosecond laser pulse excitation. We propose to explain this diversity with the notion of a tricritical point.

\begin{figure}[t!]
    \centering
    \includegraphics{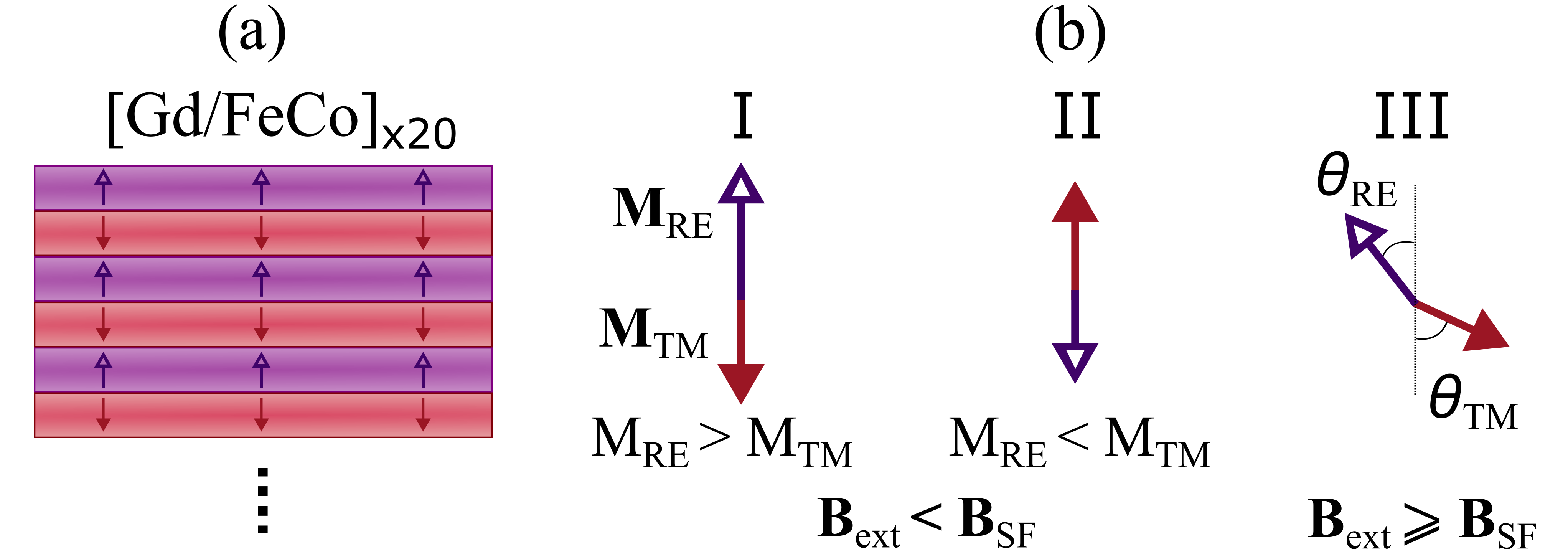}
    \caption{(a) Illustration of the spins in the multilayer. The Gd moments (purple, hollow) align antiparallel to those of the FeCo layers (red, filled). (b) The arrows depict the sublattice magnetizations of Gd and FeCo in the three static phases: two collinear (I and II) and one noncollinear phase (III).}
    \label{fig:1}
\end{figure}

The studied multilayer structure Glass/Ta($4$)/Al($10$)/ Ta($4$)/[Fe$_{87.5}$Co$_{12.5}$($0.5$)/Gd($0.5$)]$_{\times 20}$/Ta($4$), with layer thicknesses (in nm) indicated between parentheses, was grown by magnetron sputtering at an Ar pressure of $10^{-2}$~mbar. Further details on the growth procedure can be found in Ref.~\cite{https://doi.org/10.1002/admi.202201283}. The atomic magnetic moments (to which we simply refer as ``spins'') in the ferromagnetic FeCo layers with total magnetization $\mathbf{M}_{\mathrm{TM}}$ couple antiferromagnetically to the spins in the ferromagnetic Gd layers with magnetization $\mathbf{M}_{\mathrm{RE}}$, as is depicted schematically in Fig.~\ref{fig:1}(a). In the absence of an external magnetic field, the spins align along the out-of-plane anisotropy axis. Three magnetostatic phases can be distinguished in this material (see Fig.~\ref{fig:1}(b)). For low temperatures, the absolute Gd magnetization $M_{\mathrm{RE}}$ exceeds that of the FeCo layers $M_{\mathrm{TM}}$ (Phase I). As $M_{\mathrm{RE}}(T)$ decreases faster with increasing temperature than $M_{\mathrm{TM}}(T)$, a compensation temperature $T_{\mathrm{M}}$ exists at which $M_{\mathrm{TM}} = M_{\mathrm{RE}}$, and above which $M_{\mathrm{TM}} > M_{\mathrm{RE}}$ (phase II). Moreover, when an external magnetic field $\mathbf{B}_{\mathrm{ext}}$ is applied along the anisotropy axis, a critical \textit{spin-flop} field $B_{\mathrm{SF}}$ can be reached at which the gain in Zeeman energy due to an increase of net magnetization $\mathbf{M} = \mathbf{M}_{\mathrm{RE}} + \mathbf{M}_{\mathrm{TM}}$ when the spins would arrange noncollinearly exceeds its associated loss in exchange and anisotropy energy. In this case, the noncollinear phase (phase III) is attained, described by the polar angles $\theta_{\mathrm{TM}}$ and $\theta_{\mathrm{RE}}$ of the sublattice magnetizations. This spin-flop phase-transition (PT) from phases I/II to phase III can be either of the first, or second-order \cite{PhysRevB.100.064409}. In terms of Landau’s theory of PTs, in a first-order PT, the changes of the order parameter ($\theta_{\mathrm{TM}}$) are discontinuous and the transition between the phases often exhibits hysteresis. In a second-order PT, the order parameter changes continuously and without hysteresis~\cite{PhysRevB.100.064409, PhysRevB.100.174427, PhysRevApplied.13.024064}.

\begin{figure}[t!]
    \centering
    \includegraphics{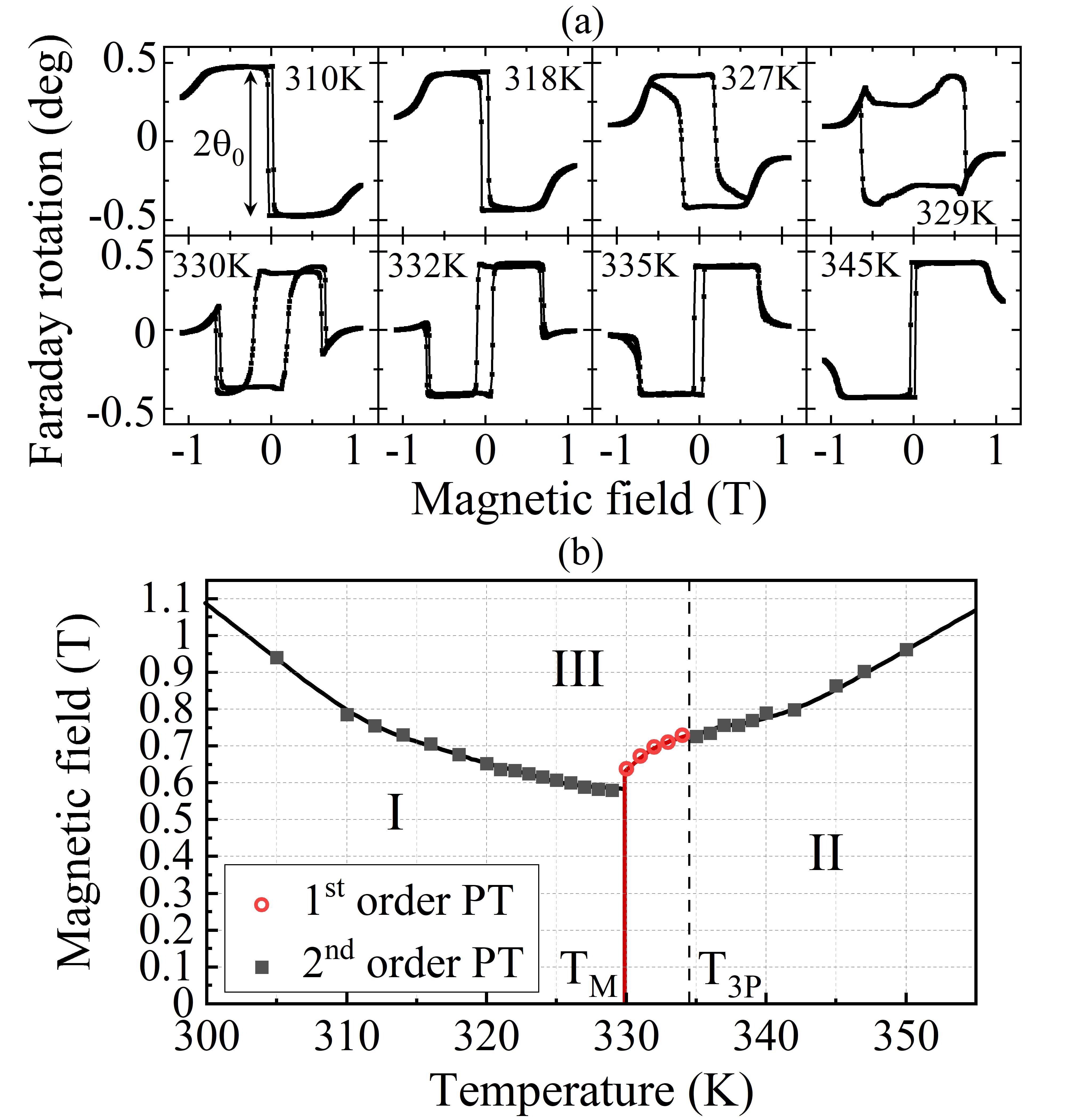}
    \caption{(a) Faraday rotation measurements of light at the wavelength $\lambda = 632.8$~nm. The bending and discontinuities of the loops correspond to the spin-flop transition.  (b) Critical field $B_{\mathrm{SF}}$ and order of the spin-flop transition deduced from the isothermal susceptibility $\chi_T$.  }
    \label{fig:2}
\end{figure}

To pinpoint $T_\mathrm{M}$ and $B_{\mathrm{SF}}(T)$ experimentally, we measured the magneto-optical Faraday effect of Gd/FeCo 
using linearly polarized light from a HeNe laser ($\lambda = 632.8$~nm) that transmitted perpendicularly through the multilayer. As commonly accepted, we assume that the Faraday rotation in GdFeCo at the wavelengths $600$-$800$~nm is dominated by the out-of-plane projection of the TM magnetization $M_{\mathrm{TM},z} = -|\mathbf{M}_{\mathrm{TM}}|\cos\theta_{\mathrm{TM}}$, and we will take this assumption throughout the article~\cite{PhysRevLett.99.217204, PhysRevB.87.180406}. The magneto-optical dominance of $\mathbf{M}_{\mathrm{TM}}$ is corroborated by the fact that the magneto-optical signal does not decrease/disappear at the magnetization compensation point and moreover remains unchanged in the temperature range $293-350$~K (see Supplemental Material \cite{Note1}). A selection of hysteresis loops measured using external magnetic fields up to $\sim 1$~T applied along the sample normal is shown in Fig.~\ref{fig:2}(a). The step around $B_{\mathrm{ext}} = 0$ corresponds to a $180^\circ$ magnetization reversal. The diverging coercive field and the flip of the loops between $329$~K and $330$~K are clear indications of a compensation temperature $T_\mathrm{M} \approx 330$~K. Moreover, the data measured at 310 K reveal a transition from a fully saturated state, which is observed at small fields, to a state with a reduced magneto-optical signal, which is observed at larger fields i.e. above $\sim 0.6 $~T. This transition sets in abruptly, but is continuous. Thus, the corresponding change in the order parameter $\theta_{\mathrm{TM}}$ can be classified thermodynamically as a second-order PT. The critical field depends on temperature and reaches a minimum at the compensation point. When the temperature is slightly above the compensation point, the continuous change has become a discontinuous jump, while also hysteresis can be observed. These features are characteristic of a first-order PT, and it persists for about $5$~K above $T_\mathrm{M}$ after which the PT changes back to the second-order. The point with temperature $T_{\mathrm{3P}} \approx 335$~K at which the first and second-order PTs meet is by convention called a \textit{tricritical point}.
\begin{figure*}[t!]
    \centering
    \includegraphics{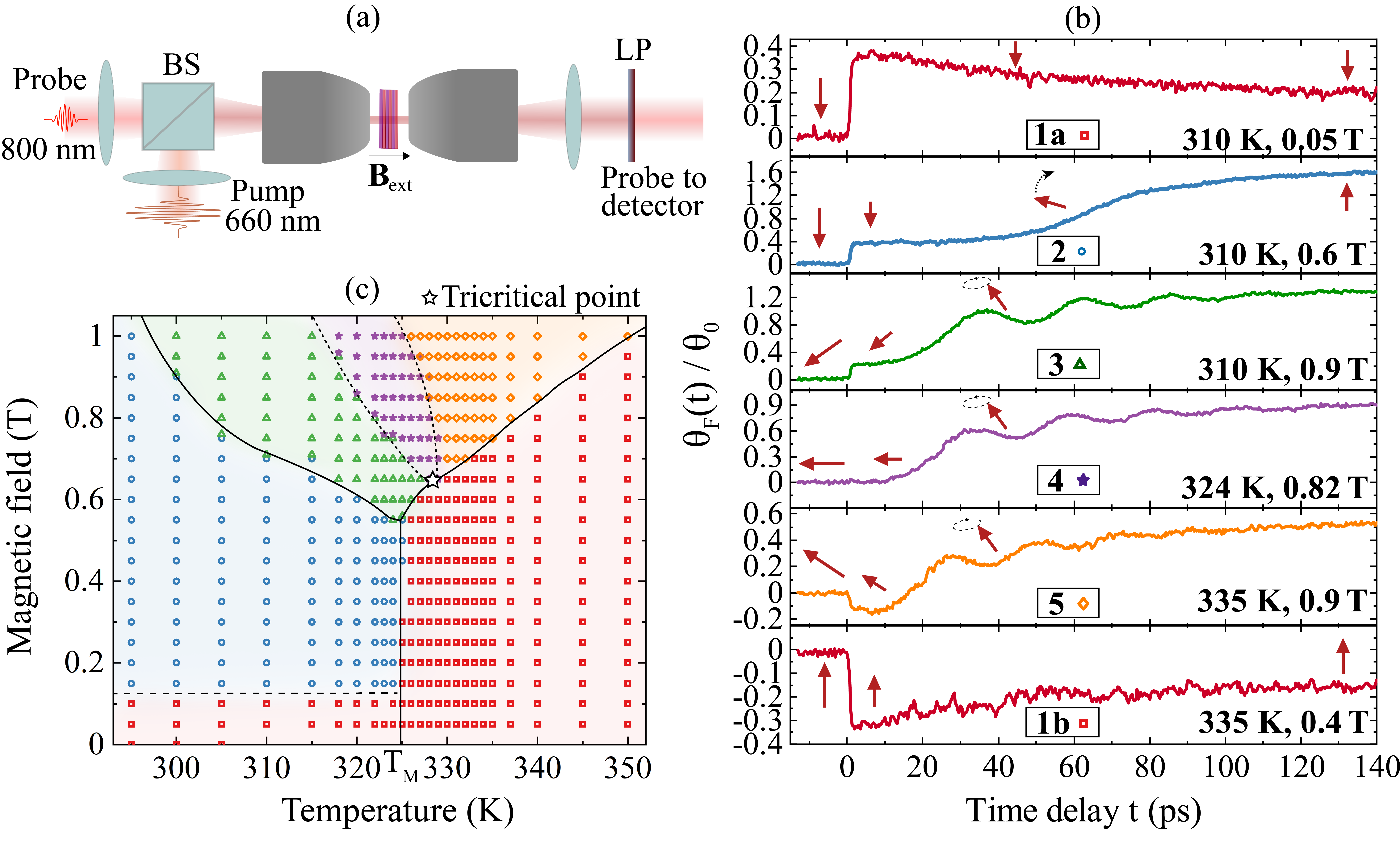}
    \caption{(a) Illustration of the experimental pump-probe setup (abbreviations: beamsplitter (BS), and long-pass filter (LP)). (b) The six different types of magnetization dynamics that were distinguished in the pump-probe measurements. The rotation data divided by $\theta_0$ can be interpreted as $\theta_\mathrm{F}(t) / \theta_0 = \Delta M_{\mathrm{TM, z}}(t)/ |M_{\mathrm{TM}}(0)|$. The arrows provide a pictorial description of $\mathbf{M}_{\mathrm{TM}}$, based on our interpretation of the data. (c) Overview of where each type of dynamics (see (b)) occurs in a $H$-$T$ diagram. Note that $T_\mathrm{M}$ shifted $\sim 5$~K compared to the results in Fig.~\ref{fig:2} due to heat accumulation in the sample. The purple area denotes the region where the amount of UD decreased below $20\%$ of the maximal demagnetization measured at low fields ($B_{\mathrm{ext}} < 0.5$~T), where the system started in the collinear state. It can be seen that the purple region terminates near the tricritical point.  }
    \label{fig:3}
\end{figure*}

The exact locations of the second-order PTs were determined by taking the center of a sigmoid function that was fitted to the isothermal susceptibility $\chi_T = \left(\frac{\partial M_{\mathrm{TM}}}{\partial B}\right)_T$. Analogously, a first-order PT is marked by a divergence of this susceptibility. The resulting critical field $B_{\mathrm{SF}} $ as a function of $T$ is summarized in Fig.~\ref{fig:2}(b). The temperature dependence of the critical field is in perfect agreement with the spin-flop transition which has been observed and predicted for GdFeCo \cite{PhysRevLett.118.117203, PhysRevB.100.064409} and TbFeCo \cite{PhysRevB.100.174427} alloys, except that the typically required field decreased nearly an order of magnitude. The results are supported by measurements using a SQUID-VSM, which probes net magnetization $\mathbf{M}$ instead of $M_{\mathrm{TM},z}$ (see Supplemental Material~\footnote{\label{footnote}See Supplemental Material at [URL] for the full static magnetic characterization of the Gd/FeCo sample using magneto-optics and the SQUID-VSM, details of the optical pump-probe experimental method, all of the measured dynamical transients, estimations of the temperature increase due to laser-induced heating, measurements of UD with $100$~fs time-resolution and an example of fluence dependence of the data, which includes Refs.~\cite{kools2023aging, PhysRevB.84.024407}.})).  

Theoretical work showed that the tricritical point should appear either above or below $T_\mathrm{M}$ depending on whether the magnetic anisotropy of the RE sublattice is larger or smaller than that of the TM sublattice \cite{PhysRevB.100.064409}. Yet, experimental studies of the noncollinear state in GdFeCo and TbFeCo alloys have not revealed the role of the tricritical point, due to the experimentally challenging requirement of high magnetic fields. Having manufactured a Gd/FeCo multilayer whose tricritical point appears well below $1$~T, we could study the laser-induced spin dynamics in detail using a table-top optical pump-probe setup with external field applied out of the sample plane as depicted schematically in Fig.~\ref{fig:3}(a) (see Supplementary Material \cite{Note1} for details of the method). The pump-induced changes of the Faraday rotation $\theta_\mathrm{F}(t)$ is again assumed to be proportional to $M_{\mathrm{TM},z}$. We divided $\theta_{\mathrm{F}}(t)$ by the static magneto-optical rotation $\theta_0$ associated with $\mathbf{M}_{\mathrm{TM}}$ pointing completely out of the plane, i.e. $2\theta_0 \approx 0.9^\circ$ equals the Faraday rotation contrast (at $\lambda = 800$~nm) in the case of a full magnetization reversal (see Fig.~\ref{fig:2}(a)). This ratio equals the normalized changes of the out-of-plane projection of the FeCo magnetization $\Delta M_{\mathrm{TM, z}}(t)/ |\mathbf{M}_{\mathrm{TM}}(0)| = \theta_\mathrm{F}(t) / \theta_0 $. For example, $\theta_{\mathrm{F}}(t) / \theta_0 = 2$ implies a full $180^\circ$ reversal of $\mathbf{M}_{\mathrm{TM}}$. 
\begin{figure}[t!]
    \centering
    \includegraphics{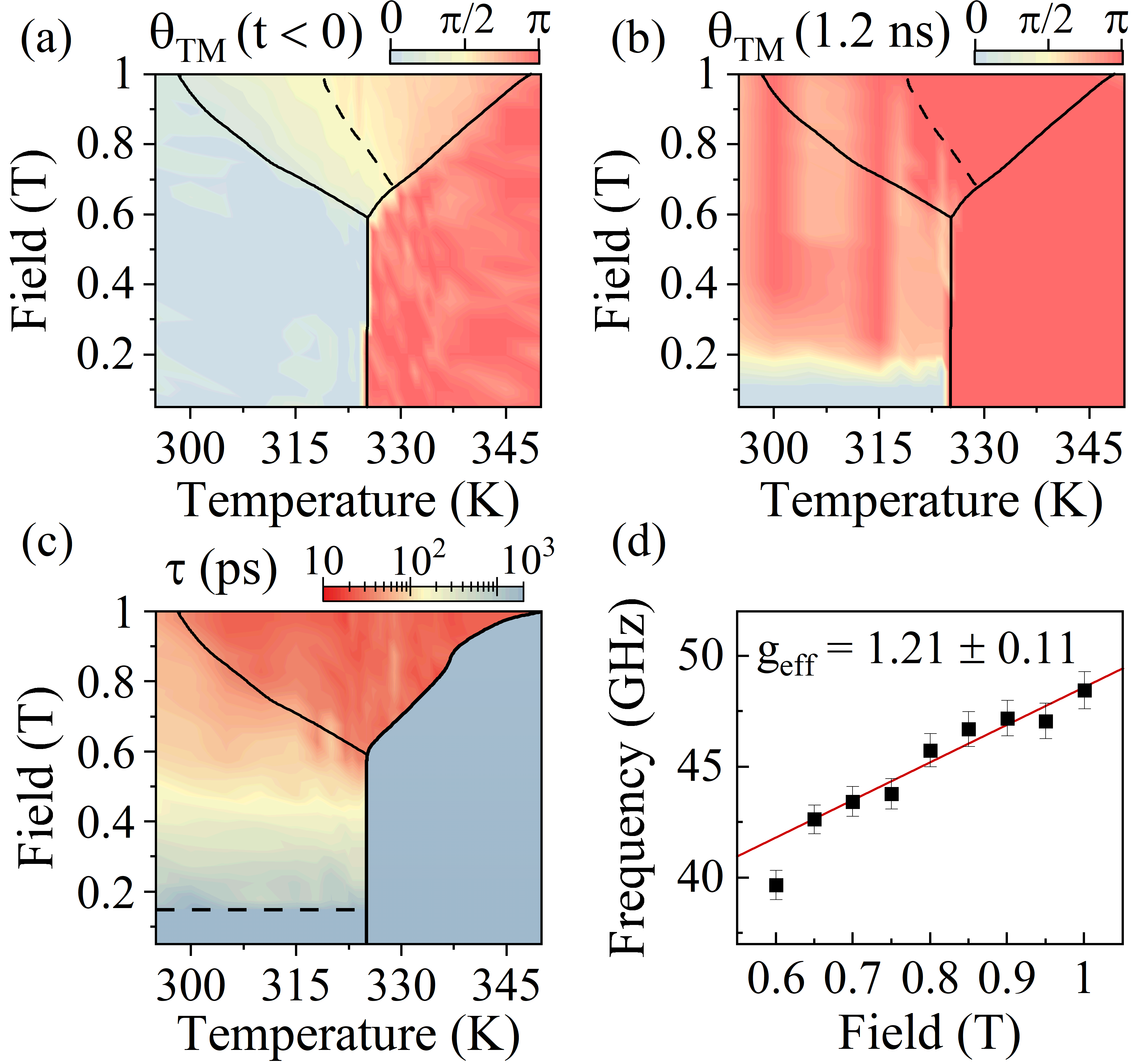}
    \caption{Several quantities derived from the pump-probe measurements: (a) The initial polar angle of the FeCo magnetization $\theta_{\mathrm{TM}}(t<0)$. The dotted line indicates where $\theta_{\mathrm{TM}} = \pi/2$. (b) The final value of $\theta_{\mathrm{TM}}$ after $1.2$~ns. The outcome is prone to several measurement and calibration uncertainties. (c) The typical timescale associated with the sub-ns dynamics of $\mathbf{M}_{\mathrm{TM}}(t)$. No reversal was observed below the dotted line on the left and in the blue region on the right. (d) Frequencies of the oscillations measured at $T = 325$~K as a function of the external magnetic field. The slope can be related to the effective $g$-factor, excluding the \textit{soft mode}~\cite{Venkataraman1979} near the spin-flop PT at $B_{\mathrm{ext}} = 0.6$~T.}
    \label{fig:4}
\end{figure}

Using this method, we observed a remarkable diversity in magnetization dynamics for different external fields and temperatures. In particular, we were able to distinguish six different types of dynamics depicted in Fig.~\ref{fig:3}(b). The curves of type $1$a and $1$b correspond to a sub-ps laser-induced partial loss of magnetic order, known as ultrafast demagnetization (UD)~\cite{PhysRevLett.76.4250}, of about $\sim 30~\%$ followed by a recovery of the FeCo-sublattice, measured below and above the magnetization compensation point for the Gd-dominated and FeCo-dominated phases I and II, respectively. Type $2$ can be assigned to UD of FeCo followed by a slow (sub-ns) reversal of the magnetization. The dynamics of types $3$ and $5$ have a reduced demagnetization compared to the types $1$ and $2$, followed by an ultrafast reorientation of the spins accompanied by spin oscillations. Type $3$ and $5$ could only be observed for high fields ($>0.5$~T), therefore we attribute the decrease in demagnetization to the reduced $z$-projection ($\propto \cos \theta_{\mathrm{TM}}$) of the magnetization in a noncollinear configuration. Also, the UD has an opposite sign in type $3$ and $5$, which indicates that the $\mathbf{M}_{\mathrm{TM}}$  pointed originally below and above the sample plane, i.e. $\theta_{\mathrm{TM}} \lessgtr \frac{\pi}{2}$, respectively (where ``above'' points in the direction of the external magnetic field). Naturally, an intermediate regime of dynamics (type $4$) exists which is similar to $3$ and $5$, but without observable UD. We assigned type $4$ when the UD decreased below the arbitrarily chosen border of $<20~\%$ of the original degree of UD (in type $1$ and $2$), in which case the degree of UD is comparable to the experimental noise.

The field and temperature where each type of spin dynamics was observed (the \textit{dynamical $H$-$T$ phase diagram}) are shown in Fig.~\ref{fig:3}(c). All the raw curves associated with each point are plotted in the Supplemental Material~\cite{Note1}. We compared this dynamical $H$-$T$ phase diagram to the static diagram of Fig.~\ref{fig:2}(b), and discuss the conclusions that can be drawn from that. To aid this discussion, we calculated several quantities from the dynamical measurements shown in Fig.~\ref{fig:4}. The calculation of the initial polar angle of the FeCo magnetization $\theta_\mathrm{TM}(t<0)$ in Fig.~\ref{fig:4}(a) was based on the reduced out-of-plane projection of UD in the noncollinear phase as compared to the collinear phase (see Supplemental Material~\cite{Note1} for details). Using these initial angles, we determined the final state angle at $1.2$~ns as is depicted in Fig.~\ref{fig:4}(b). In addition, we fitted the growth of the signal right after UD with a sigmoidal function and extracted a timescale in which $\sim96\%$ of the growth takes place (see Supplementary information). The results are depicted in Fig.~\ref{fig:4}(c).

By comparing the dynamical phase-diagram of Fig.~\ref{fig:3}(c) with the results of Fig.~\ref{fig:4}, we first conclude that the flip in the sign of the UD between types $1$a and $1$b marks the compensation temperature $T_\mathrm{M}$, which has shifted with about $5$~K compared to the static phase diagram likely due to heat accumulation in the sample. Furthermore, for $T < T_{\mathrm{M}}$ and for fields $0.15$~T $\leq B_{\mathrm{ext}} < B_{\mathrm{SF}}$ (type $2$), the spins undergo a sub-ns magnetization reversal after the UD. The end-state after $1.2$~ns is independent of the applied external field (see Fig.~\ref{fig:4}(b)), which implies that FeCo undergoes a full reversal to phase II. An exception might be the results just above the threshold field of $0.15$~T, seen as the yellow line in Fig.~\ref{fig:4}(b). Here, the final state could correspond to a non-collinear configuration, but it might also correspond simply to a partially reversed spot. In Fig.~\ref{fig:4}(c), one can see that above the threshold field of $0.15$~T (see dashed line), the timescale of reversal gradually reduces for larger external fields from $\sim 1$~ns to about $100$~ps upon approaching phase III. We estimated that the sample temperature is elevated by laser-heating to about $140$~K after $1.2$~ns (see Supplemental Material~\cite{Note1}). Therefore, these observations let us conclude that reversal of type $2$ occurs via transverse motion after the system is laser-heated above the compensation point.

Next, we address the dynamics of types $3$-$5$, which are only triggered when the system starts in the noncollinear state (phase III). Figure~\ref{fig:4}(b) shows that the final angle in this entire region reaches $\theta_{\mathrm{TM}} \sim \pi$, suggesting that after laser-induced heating the system ends up in phase II. Most interestingly, the typical timescale of this spin reorientation is $\sim 20$~ps, which is about an order of magnitude faster than the spin reversal starting from the collinear phase I. This ultrafast timescale is roughly independent of the applied magnetic field (see Fig.~\ref{fig:4}(c)), indicating it is exchange-driven. One should note, however, that despite the reorientation being fast, it is followed by spin precession around the new equilibrium and the magnetization will eventually arrive at this new equilibrium only after the oscillations are damped. The frequency of precession $\nu_{\mathrm{SF}}$ is faster than expected for the ferromagnetic resonance mode in a collinear state \cite{PhysRevB.73.220402, hintermayr2023exploring}, hence this mode should correspond to the \textit{spin-flop resonance} \cite{gurevich1996magnetization, PhysRevLett.118.117203}. The mode frequency follows a slope defined by the \textit{effective $g$-factor}, which is found to be $1.21\pm 0.11$ near $T_M$ (see Fig.~\ref{fig:4}(d)). Similar values of $g_{\mathrm{eff}}$ near $T_M$ of the spin-flop resonance mode in GdFeCo were reported in high fields~\cite{PhysRevLett.118.117203}. 

Finally, the other key observation in our work regards the novel insight about the tricritical point, as can be deduced from the dynamics in the purple region (type $4$ in Fig.~\ref{fig:3}(c)), where the demagnetization is nearly invisible or completely absent and therefore corresponds to the situation where the $\mathbf{M}_{\mathrm{TM}}$  started in the plane $\theta_{\mathrm{TM}} \approx \pi/2$. This region is seen to terminate at the spin-flop transition, exactly where the tricritical point is expected. This allows us to establish a novel empirical rule, which states that the tricritical point separates the regions with $\theta_{\mathrm{TM}} \lessgtr \frac{\pi}{2}$, i.e. when $\mathbf{M}_{\mathrm{TM}}$ initially points towards different sides of the anisotropy plane. This rule can be explained intuitively: when initially $\theta_{\mathrm{TM}} < \frac{\pi}{2}$, $\mathbf{M}_{\mathrm{TM}}$ has to jump over a potential barrier at $\theta_{\mathrm{TM}} = \frac{\pi}{2}$ due to magnetic anisotropy to reach the TM-dominated collinear phase II (where $\theta_{\mathrm{TM}} = \pi$). This jump is discontinuous and can thus be classified as a first-order PT. Naturally, when $T<T_{\mathrm{M}}$ or when $T > T_\mathrm{3P}$, $\mathbf{M}_{\mathrm{TM}}$ remains to point to the same side of the plane during the spin-flop transition. Therefore it has no potential barrier to pass, and the associated spin-flop PT can be continuous and thus of the second order.

In conclusion, our results highlight the unique diversity and tunability of magnetization dynamics in RE-TM ferrimagnets, and the established phase diagram provides new insights into the role and origin of the so far illusive tricritical point present in these materials. The extensive dataset provides many parameters associated with the statics and dynamics of the noncollinear phase, which could not be resolved before. We note that despite intense studies of ultrafast magnetism in ferrimagnets during the last 15 years, the role of the tricritical point in the diversity of the spin dynamics has not been treated either theoretically or computationally. Finally, our results demonstrate the possibility of switching between the noncollinear and collinear spin configurations on ultrafast $\mathcal{O}(10~\text{ps})$ timescales. This puts the non-collinear phase forward as a potentially interesting host of bits in magnetic recording.

\begin{acknowledgments}
The authors thank K. Saaedi, C. Berkhout, and S. Semin for their technical support. The work was supported by the European Research Council ERC Grant Agreement No.101054664 273 (SPARTACUS), and the research program ``Materials for the Quantum Age'' (QuMat). This program (registration number 024.005.006) is part of the Gravitation program financed by the Dutch Ministry of Education, Culture and Science (OCW).

\end{acknowledgments}


\bibliography{references}

\end{document}


\title{Supplemental Material for ``Diversity of Ultrafast Spin Dynamics Near the Tricritical Point 
in a Ferrimagnetic Gd/FeCo Multilayer''}

\date{\small{\today}}

\maketitle
\textbf{Table of contents}

\tableofcontents 
\newpage
\section{Static characterization with a SQUID-VSM}
\begin{figure}[h!]
    \centering
    \includegraphics[scale = 0.9]{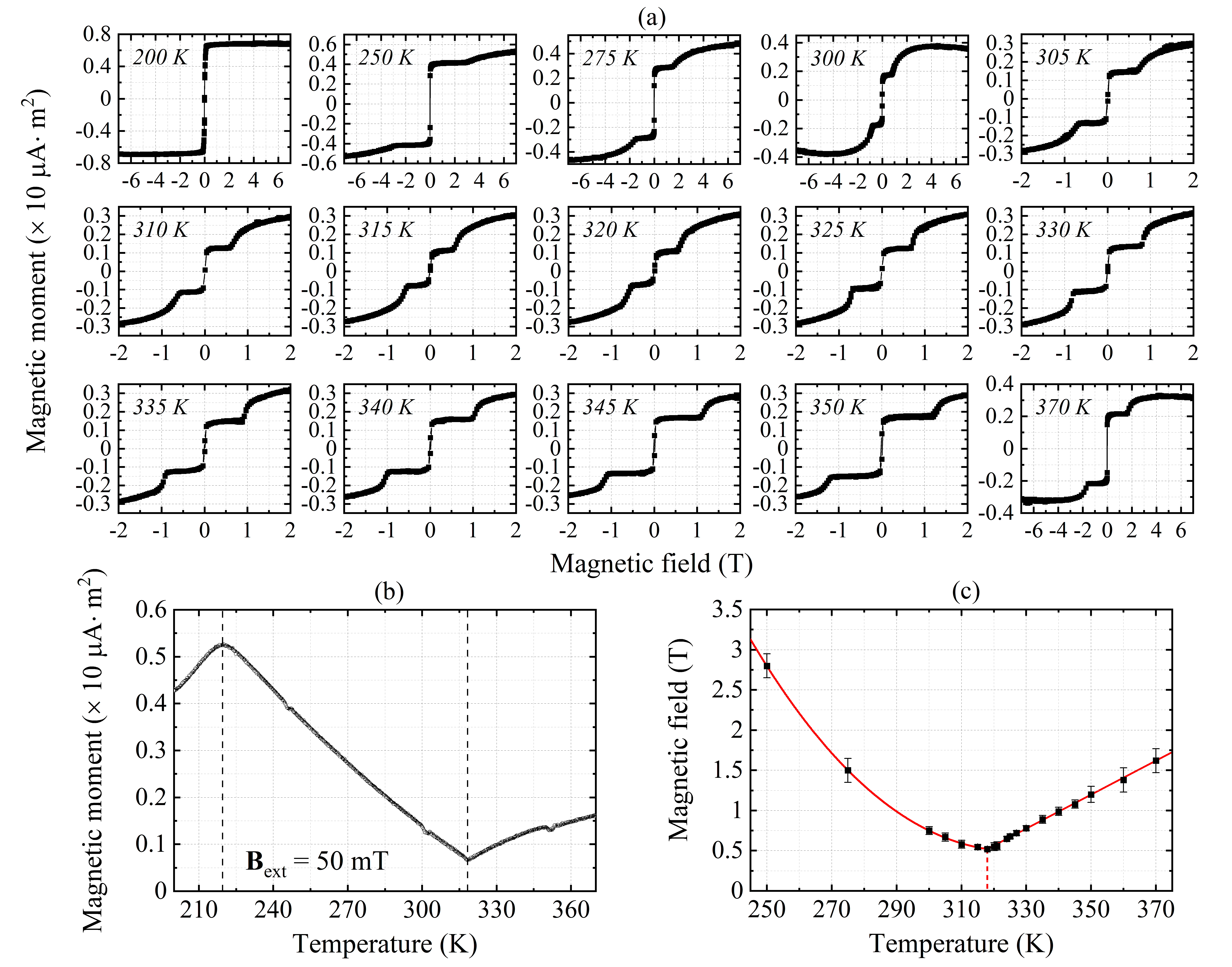}
    \caption{\small{(a) Measurements of the net magnetic moment of the Gd/FeCo along the sample normal using a VSM-SQUID, where the external field was applied along the same normal. Note that the Gd/FeCo sample studied here was grown on SiO, as opposed to the sample considered in the rest of the article which was grown on glass. (b) Net magnetic moment as a function of temperature, measured with an applied field of $50$~mT. The field was applied to create a favorable orientation of the net magnetic moment. The dip at $320$~K corresponds to the magnetization compensation temperature, while the discontinuity at the slope at $\approx 220$~K corresponds to a loss of out-of-plane anisotropy of the sample. Such loss occurs when the temperature is far from the compensation point, in which case the net magnetization becomes so large that the demagnetization field (also known as shape anisotropy) pulls the magnetization in the plane. (c) The location of the spin-flop transition is determined manually from the hysteresis loops, where the error bars depict the estimated error of this manual determination.} } 
    \label{fig:SQUID}
\end{figure}
In Fig.~\ref{fig:SQUID}(a), the step around the zero fields corresponds to magnetization reversal, and this step reaches a minimum at $T \approx 320$~K. This temperature marks the compensation temperature. Note that the compensation temperature is therefore shifted with about $10$~K compared to the results obtained by the magneto-optical Faraday effect in the coming section \ref{sec:Faraday}. This shift can be attributed to several factors: the use of a different substrate, calibration uncertainties, and different thermal coupling among devices. Moreover, the time between these measurements and the Faraday effect measurements (see Sec.~\ref{sec:Faraday}) was about $1$~year. This could influence the compensation point, as the interlayer diffusion is known to change over time~\cite{kools2023aging}.
\clearpage

\section{Static characterization using the magneto-optical Faraday effect\label{sec:Faraday}}
The selection of the following data was presented in Fig.~$2$ of the main article, here we plot the full set of calibrated hysteresis loops and discuss several more details. 
\begin{figure}[b!]
    \centering
    \includegraphics{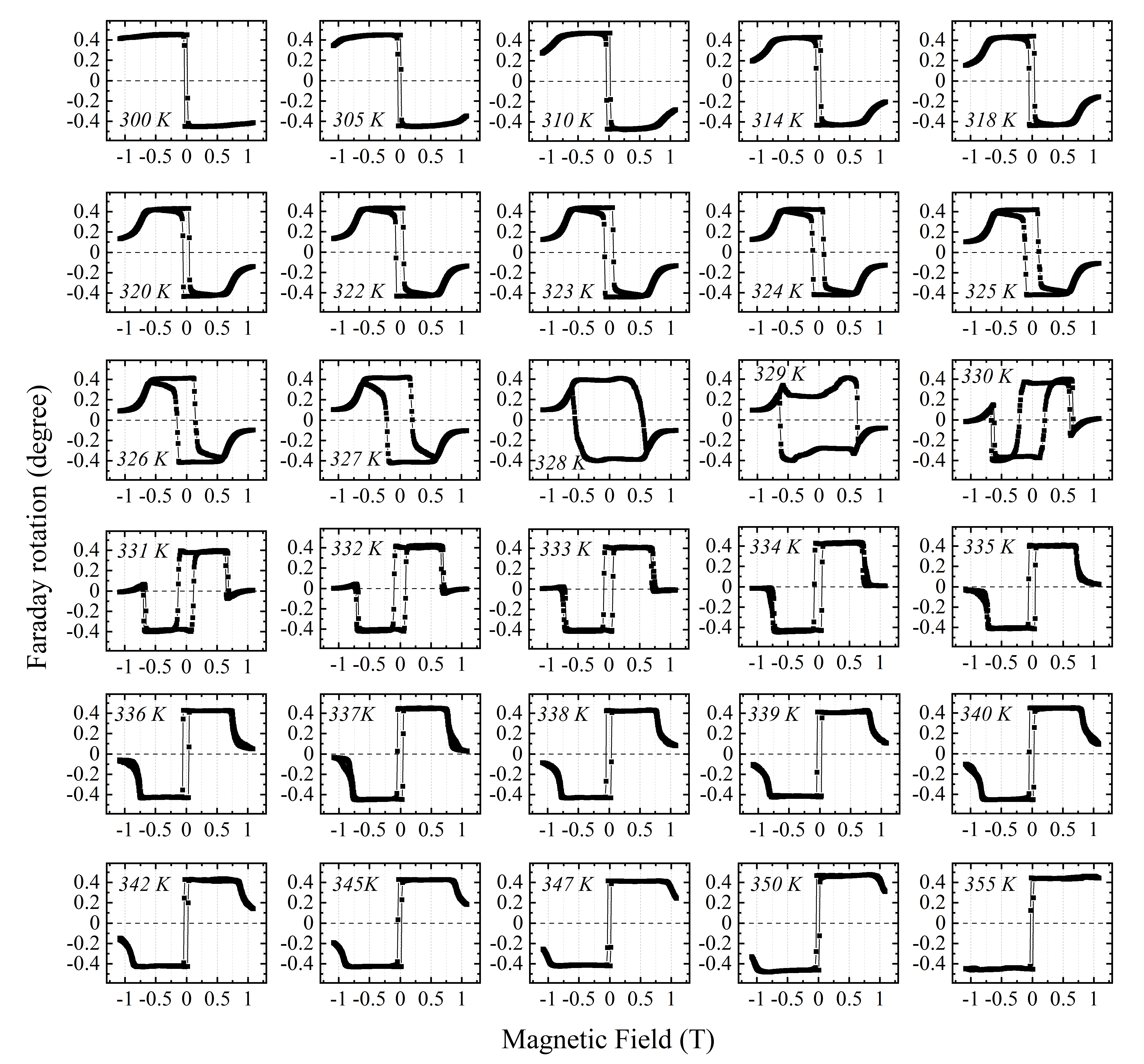}
    \caption{Magneto-optical Faraday rotation as a function of the external magnetic field of light origination from a HeNe laser ($\lambda = 632.8$~nm), measured at various around the compensation point.}
    \label{fig:HeNe}
\end{figure}

It can be seen that the step around $B_{\mathrm{ext}} = 0$ does not drop at the compensation point, which implies that one of the two magnetic sublattices dominates the magneto-optical signal. That is, if both sublattices contributed comparably, then the loop should drop at the compensation point (as was seen in the SQUID-VSM measurements of the previous section, where the net magnetization is probed). Moreover, the fact that the steps of the loops in the entire temperature range $300$-$355$~K are nearly unchanged, agrees with the assumption that the FeCo dominates the magneto-optical signal, as $M_{\mathrm{Gd}}(T)$ should depend strongly on the temperature in this range. Therefore, if $M_{\mathrm{Gd}}$ would have been dominated, an appreciable dependence of the Faraday effect on temperature is expected, which is not the case.

Note that one of the main conclusions of the article, which states that the tricritical point marks the separation between the regions where the FeCo magnetization on adjacent sides of the sample plane, is also subtly visible in this data: the first-order spin-flop transition (discontinuous jump, hysteresis) only occurs when the Faraday signal reaches below the dotted zero line. According to the interpretation of the Faraday signal being proportional to the out-of-plane component of the FeCo magnetization, this line can only be crossed if the FeCo magnetization reaches the opposite side of the plane. This means that we see a first-order phase transition if and only if the FeCo magnetization goes to the adjacent side of the plane (opposite to where the field is pointing) during the spin-flop transition, exactly what we concluded from the dynamical data. During a second-order spin-flop phase transition, the FeCo magnetization remains on the same side of the plane.
\clearpage

\section{Calibrated time-resolved pump-probe measurements}
In this section, we depict all of the calibrated waveforms that were recorded during the pump-probe measurements and which have been used to produce Figures $3$ and $4$ of the main article.
\label{sec:rawdata}
\subsection{Methods}
We employed amplified linearly polarized pump and probe pulses with $~100$~fs pulse duration at central wavelengths of $\lambda = 660$~nm and $\lambda = 800~$nm, respectively. The repetition rate of the pulses was $1$~kHz, but the pump pulses were modulated at half the repetition rate using a mechanical chopper for the purpose of measuring with a lock-in amplifier. The pulses were brought to the same optical path using a beamsplitter (BS) and were normally incident and focused in the same spot on the sample. The pump beam was focused with a beam radius of $w \approx 400$~$\mu$m, and the probe pulses were focused tightly within this spot with $w \approx 70$~$\mu$m. The pump pulse energy was kept constant at $E_p = 2.8 \pm 0.3$~$\mu$J, corresponding to a peak fluence $F = E_p/(\pi w^2 /2)$ of approximately $1.1 \pm 0.2$~mJ/cm$^2$. A uniform external magnetic field up to $1$~T was applied normally to the sample, along the easy axis of magnetic anisotropy. After the pulses were transmitted through the sample, we blocked the pump pulses using a long pass (LP) filter. The probe pulses were collected by a balanced photodetector which measured the change of polarization rotation as a function of the pump-probe delay time $t$ with the help of a lock-in amplifier. Practically all dynamics were recorded both on ``short'' ($<140$~ps) and ``long'' ($1.2$~ns) timescales.
\newpage 

\subsection{Time-resolved data}
\begin{figure}[h!]
    \centering
    \includegraphics{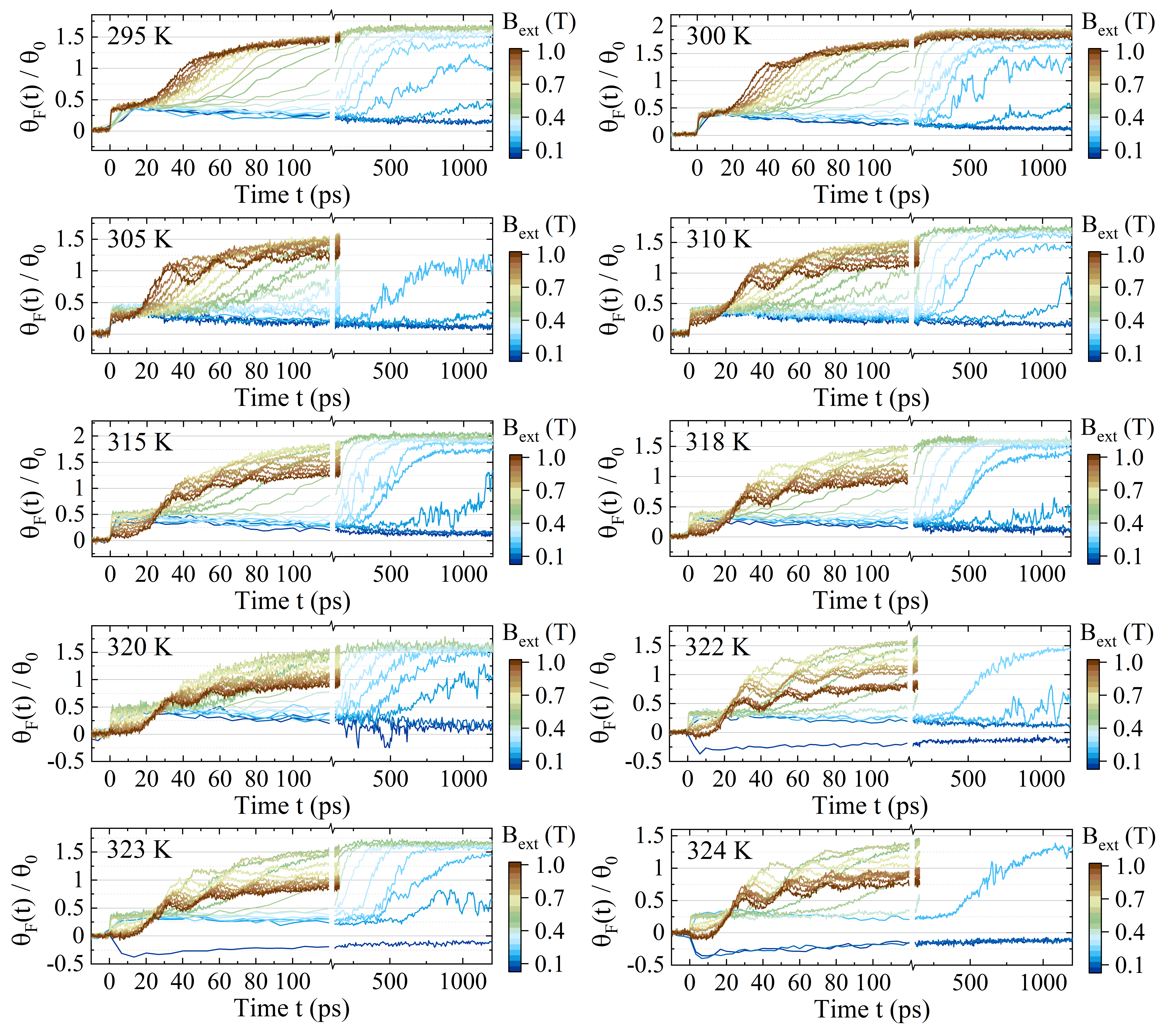}
    \caption{Waveforms measured in the range $295$~K to $324$~K for various external magnetic fields. The data has been calibrated using the step size in probe polarization rotation $2\theta_0\approx 0.9^\circ$ around $B_{\mathrm{ext}} = 0$~T of the hysteresis loops that were recorded alongside the pump-probe dynamical measurements. One should be able to distinguish all types of spin dynamics as discussed in the main article. Note that the fact that the waveforms of type $2$ (where the system starts in the Gd-dominated collinear state (II)) saturated to the same value suggested that the spins undergo a full reversal, as otherwise field dependence of this reversal is expected. This is also confirmed by the results presented in Fig.~$4$ of the main article. However, special attention should be paid to the waveforms for fields around $\sim 0.15 - 0.2$~T, which are shown to saturate at a lower value than the other curves of type $2$, indicating that the spins might have ended in a noncollinear configuration.   }
    \label{fig:295K-324K}
\end{figure}
\clearpage

\begin{figure}[h!]
    \centering
    \includegraphics{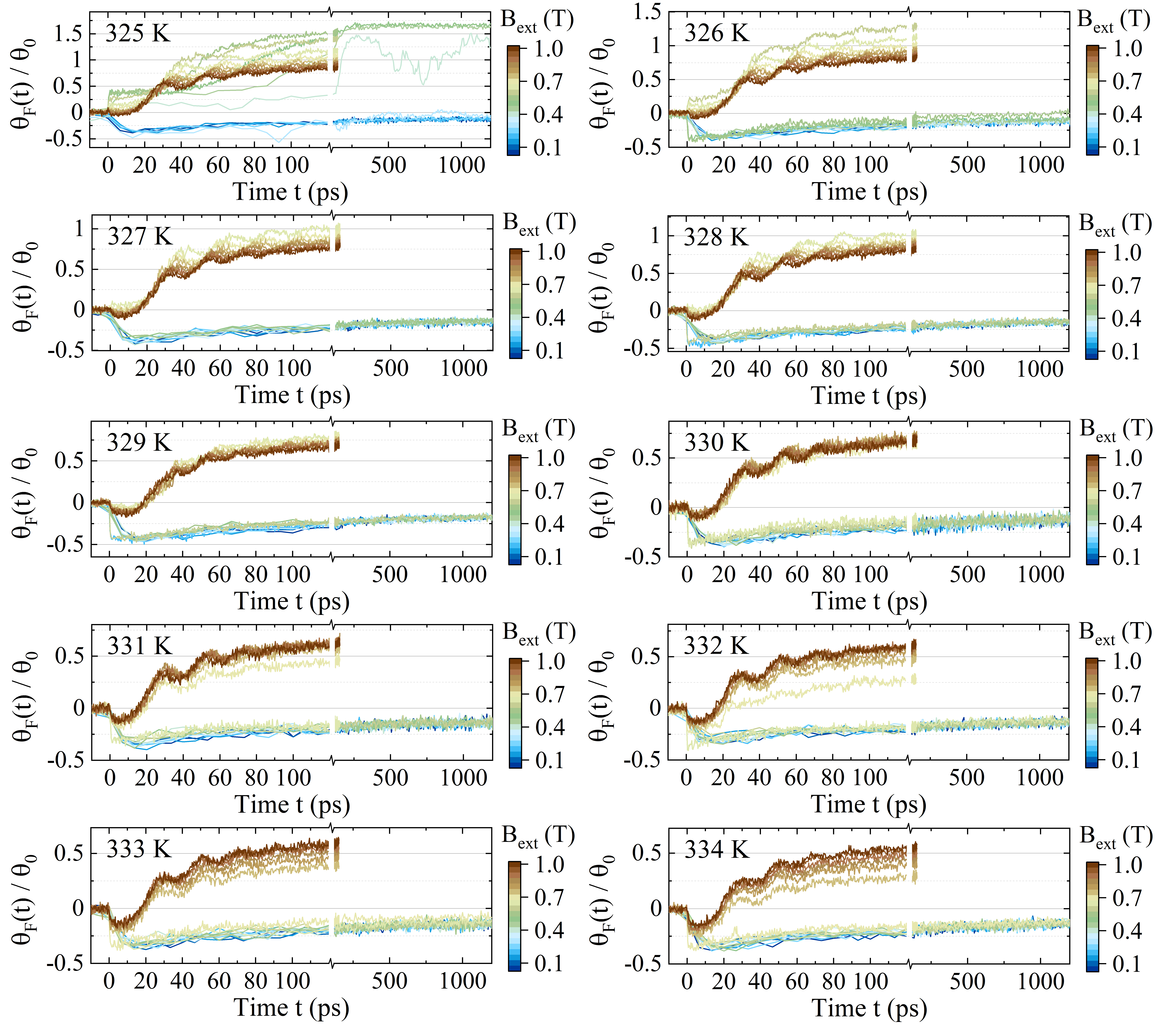}
    \caption{Waveforms measured in the range $325$~K to $334$~K for various external magnetic fields. 
    }
    \label{fig:325K-334K}
\end{figure}
\clearpage 

\begin{figure}[h!]
    \centering
    \includegraphics{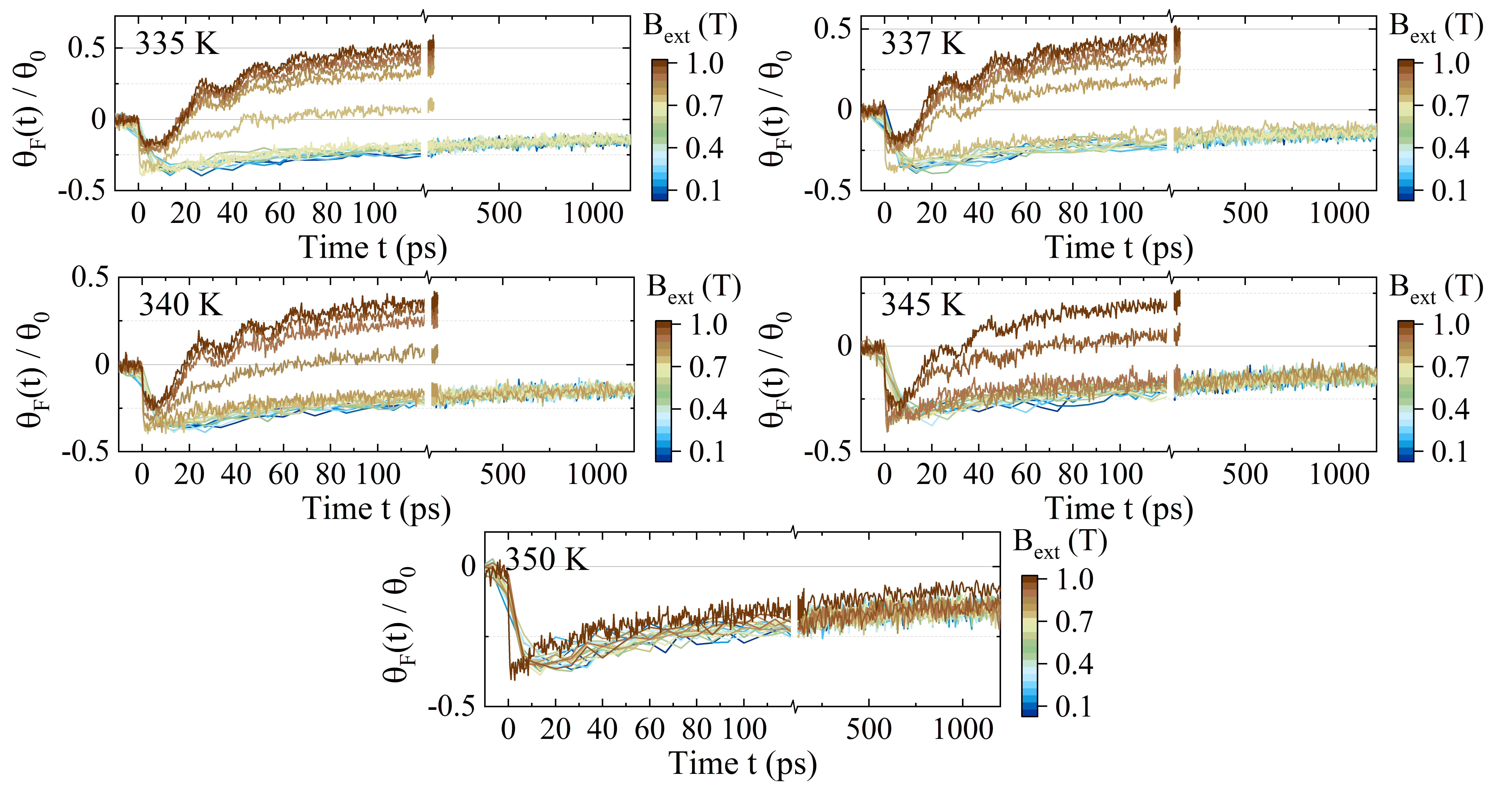}
    \caption{Waveforms measured in the range $335$~K to $350$~K for various external magnetic fields.} 
    \label{fig:335K-350K}
\end{figure}
\subsection{Procedure of deducing parameters from the data (Fig.~$4$)}
In Fig.~$4$ of the main article, the initial state $\theta_{\mathrm{TM}}(t=0)$ is derived from the reduced degree of observed demagnetization when starting in the noncollinear phase as compared to the collinear phases. This reduction is assumed to be due to the reduced projection of the $z$-component of $\mathbf{M}_{\mathrm{TM}}$ in the non-collinear state.

In particular, by comparing the amount of demagnetization $\Delta_M$ of any curve in the $H$-$T$ phase diagram with the degree of UD measured in type $1$ ($\Delta_{M0}$), one can show that $$\theta_\mathrm{TM}(t<0) = \cos^{-1}(\Delta_M/\Delta_{M0}).$$ Ultimately, we observed very weak or no demagnetization ($\Delta_M/\Delta_{M0} < 0.2$) in the dynamics of type $4$, therefore we concluded here that $\theta_{\mathrm{TM}} \approx \pi/2$, i.e. the FeCo magnetization is in the plane. Afterward, based on these initial angles, we could determine the final state at $1.2$~ns as is depicted in Fig.~$4$(b) of the main article.

The typical timescales of switching depicted in Fig.~$4$(c) of the main article are derived by fitting the sigmoidal function $$S(t) = A_1 + \frac{A_2}{1+e^{(t-t_0)/\Delta t}}$$ to the growth of the data, where $A_1$ and $A_2$, are the background signal and the amplitude, respectively, $t_0$ marks the start of the growth, and $\Delta t$ is a time parameter related to the timescale of the growth. The time at which $96\%$ of the growth takes place is $\tau = 8\Delta t$, which is the value that has been depicted in the contour plot of Fig.~$4$(c) of the main article. 
\clearpage
\section{Electron temperature heating estimation}
\subsection{Temperature increase estimated by the Curie-Weiss law}
The temperature-development of the transition-metal (FeCo) magnetization $M_{\mathrm{TM}}(T)$ follows approximately the Curie-Weiss law:
\begin{equation}
    M_{\mathrm{TM}}(T) = M_{\mathrm{TM}}\sqrt{\frac{T_{\mathrm{C}} - T}{T_{\mathrm{C}}}}
\end{equation}
where $T_\mathrm{C}\approx 800$~K is the Curie temperature. This value $T_\mathrm{C}$ is not known exactly but has been estimated from what is known for GdFeCo alloys~\cite{PhysRevB.84.024407}. In the measurements, we started at a temperature $T_0$ where the magnitude of the FeCo magnetization equals $M_{\mathrm{TM}}(T_0)$. The laser pulse heats the medium and increases the electron temperature by $\Delta T$. As a result, the magnetization decreases $M_{\mathrm{TM}}(T_0 + \Delta T)$ from which we can calculate the demagnetization ratio
\begin{equation}
    \Delta_{M} \equiv 1-\frac{M_{\mathrm{TM}}(T_0 + \Delta T)}{M_{\mathrm{TM}}(T_0)}.\label{eq:demagratio}
\end{equation}
The value of $\Delta_{M}$ for any $t$ was easily read off from the low-field demagnetization curves of Section~\ref{sec:rawdata}, as can be directly read off from the value on the $y$-axis. This, in turn, allows us to give an estimate of $\Delta T$ by rewriting Eq.~\eqref{eq:demagratio}
\begin{equation}
    \label{eq:DeltaT}
    \Delta T = (T_{\mathrm{C}} - T_0)(2\Delta_{M}-\Delta_{M}^2).
\end{equation}
When at $T_0\approx 300$~K we observed on average a demagnetization of $15\%$ after $t = 1.2$~ns when measuring at low fields $B_{\mathrm{ext}} = 0.05$ - $0.1$~T (see Fig.~\ref{fig:295K-324K}). This gives an estimation of the amount of temperature increase
\begin{equation}
    \Delta T \approx 140~\mathrm{K}.
\end{equation}

\subsection{Temperature dependence versus fluence}
\begin{figure}[h!]
    \centering
    \includegraphics{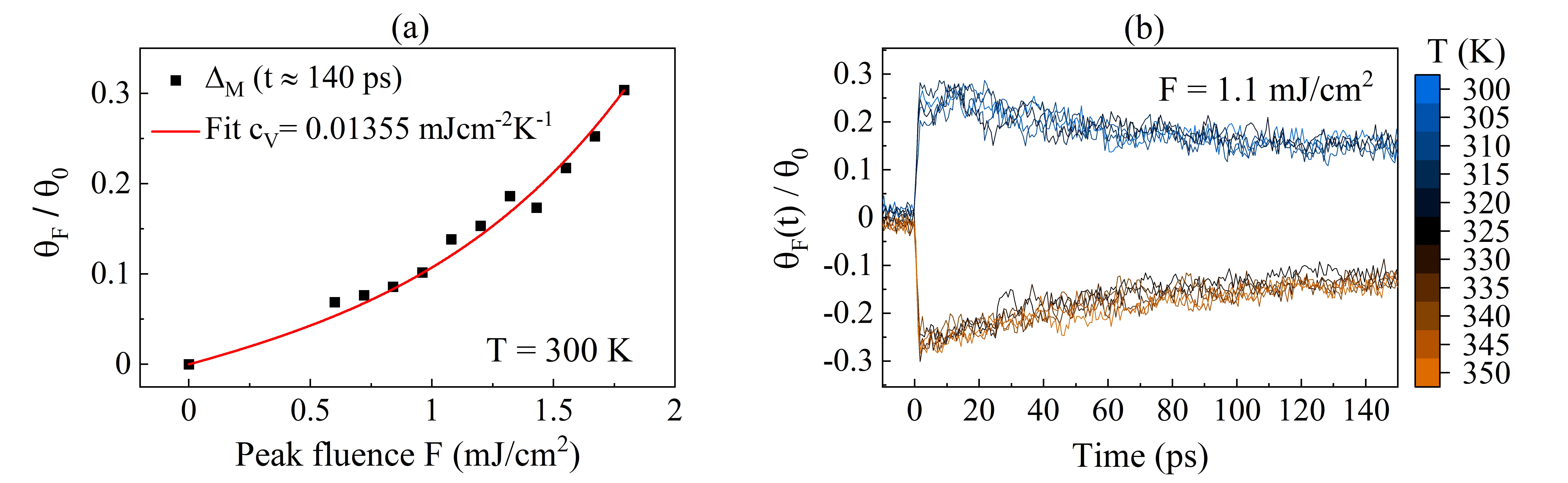}
    \caption{(a) The degree of demagnetization $\Delta_M$ measured as a function of the applied laser peak fluence $F$ at a temperature of $T = 300$~K and external magnetic field of $B_{\mathrm{ext}} = 0.05$~T. The result was fitted using Eq.~\eqref{Taylor}. (b) Ultrafast demagnetization measured at a fixed fluence and $B_{\mathrm{ext}} = 0.05$~T, but for various temperatures, showing that the degree of demagnetization is nearly independent on the temperature within the experimental temperature range between $300$~K and $350$~K.  }
    \label{fig:demag_vsF_T}
\end{figure}

The increase of temperature $\Delta T$ is proportional to the applied laser fluence:
\begin{equation}
    \label{F}
    F = c_V\Delta T
\end{equation}
where $c_V$ with units of mJcm$^{-2}$K$^{-1}$ can be interpret as heat capacity. In first-order approximation, the heat capacity is only dependent on the lattice and therefore independent of temperature. We can then rewrite Eq.~\eqref{eq:DeltaT} to obtain the demagnetization ratio $\Delta_M$ as a function of F:
\begin{eqnarray}
    \Delta_M^2 - 2\Delta_M + \frac{F}{c_V(T_C-T_0)} = 0 \implies \Delta_M = 1-\sqrt{1-\frac{F}{c_V(T_C-T_0)}}
\end{eqnarray}
and we can approximate this using the Taylor series:
\begin{equation}
    \label{Taylor}
    \Delta_M \approx \frac{1}{2c_V(T_C-T_0)}\left(F + \frac{1}{4}F^2 + \frac{1}{8}F^3 + \frac{5}{64}F^4 + \mathcal{O}(F^5) \right).
\end{equation}
In order to fit this function and find a value for $c_V$, we measured the degree of demagnetization $\Delta_M$ (after $\approx 140$~ps) as a function of $F$. The result is shown in Fig.~\ref{fig:demag_vsF_T}(a). The result was fitted using Eq.~\ref{Taylor} taking the Taylor series up to the power four ($F^4$), which gives $c_V \approx 0.01355 $~mJcm$^{-2}$K$^{-1}$. 

For the general applicability of this fitted value, we measured the demagnetization rate for constant $F$ in the entire experimental temperature range. The result shown in Fig.~\ref{fig:demag_vsF_T}(b) suggests that the degree of demagnetization is nearly identical for all temperatures within our range of interest $[300\mathrm{ K},350\mathrm{ K}]$. Therefore, the empirical law was only established for $T = 300$~K, it can be generally applied for all temperatures we considered.

\newpage
\subsection{Fluence dependence of the experimental dynamics}
In the main article, the chosen fluence was so high that nearly all transients ended in the collinear TM-dominated state (phase II). This can be rationalized using the estimated increase of temperature, which was found to be about $140$~K after $1.2$~ns: The non-collinear phase III only appears for fields smaller than $1$~T only for temperatures below $350$~K (see for example Fig.~\ref{fig:SQUID}), while $\Delta T = 140$~K implies that we heated well-above $350$~K in all measurements of Section~\ref{sec:rawdata}. Is thus obvious that the newly attained thermodynamic equilibrium will be phase II. 

\begin{figure}[h!]
    \centering
    \includegraphics{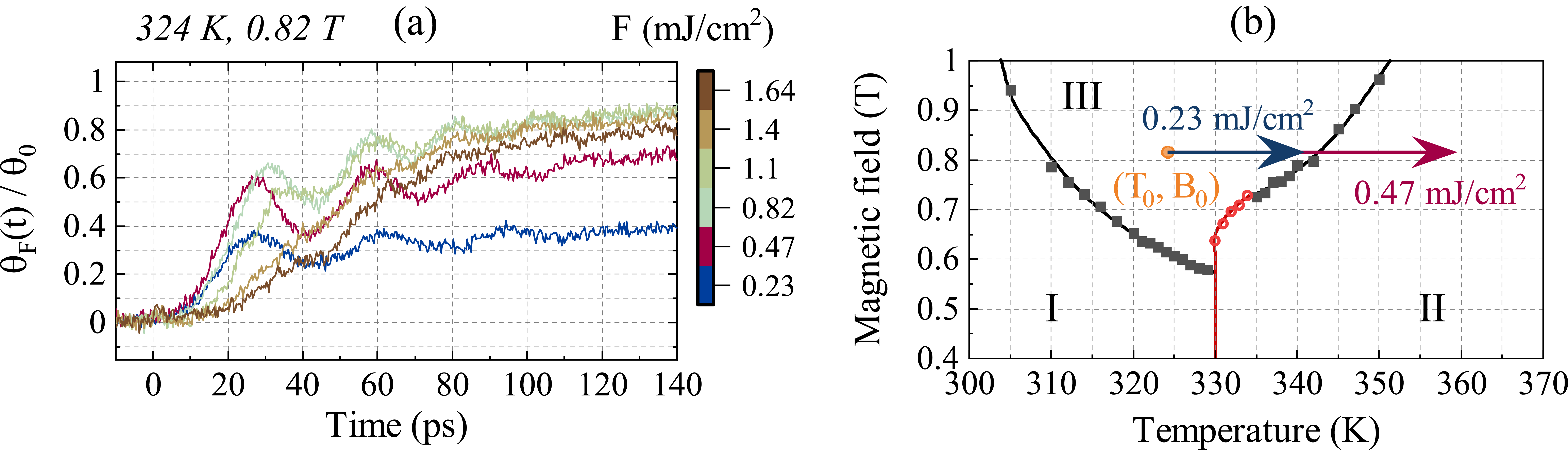}
    \caption{(a) Dynamics measured at $T_0 = 324$~K, and $B_0= 0.82$~T (type $4$) as a function of $F$, showing that the amplitude of the oscillations and the end state depend on the applied laser fluence. (b) Static phase diagram as depicted in Fig.~$2$ of the main article. The yellow dot indicates the initial conditions used in part (a). The arrows depict the estimated laser-induced increase of temperature of the lowest two fluences used in part (a), estimated using Eq.~\eqref{F} and the fitted value for $c_V$. These estimations indeed predict that for the low fluence of $0.23$~mJ/cm$^2$, the system remains in phase III, instead of undergoing a phase transition to phase II as is the case of higher fluences. This explains why the end-state signal for this lower fluence dropped significantly as compared to the higher fluences as seen in part (a).}
    \label{fig:fluence}
\end{figure}

At the same time, this means that if we reduce the applied laser-fluence $F$, a different end-state other than phase II could be reached. To this end, we measured the dynamical data of type $4$ ($T = 324$~K, $B_{\mathrm{ext}} = 0.82$~T) as a function of $F$ as shown in Fig.~\ref{fig:fluence}(a). It can be seen that the low-fluence result $F = 0.23$~mJ/cm$^2$ ends with a drastically lower signal than the other transients, indicating that phase II was not reached but rather that the spins underwent a spin reorientation within the noncollinear phase III. To show that this is consistent with the expectation for the laser-induced heating, we estimated the temperature increase $\Delta T$ after $\approx 140$~ps based on Eq~\eqref{F} using the fitted value $c_V \approx 0.01355 $~mJcm$^{-2}$K$^{-1}$. The estimated $\Delta T$ for the fluences $0.23$~mJcm$^{-2}$ and $0.47$~mJcm$^{-2}$ are $17$~K and $35$~K, respectively. We displayed the temperature increase of the two lowest fluences using arrows in the static phase diagram in Fig.~\ref{fig:fluence}(b). It can be seen that our estimation of $\Delta T$ indeed predicts that the system remains within phase III for the lowest fluence of $F = 0.23$~mJcm$^{-2}$. The spins, therefore, only undergo a spin reorientation instead of a change of phase from III to II. This result underlines that in future studies, the entire phase diagram could be studied not only as a function of field and temperature but also as a function of the applied laser fluence. For example, it can already be seen in Fig.~\ref{fig:fluence}(a) that not only the end-state depends on $F$, but also the time-scale of the growth as well as the observed amplitude of the oscillations. Exploring the dynamical phase diagram also as a function of $F$ could, therefore, lead to an even bigger variety of laser-induced spin dynamics.

\section{Ultrafast demagnetization in sub-ps time resolution}
From the data presented in the main article, it is not immediately evident if the time scale of the ultrafast demagnetization step depends on the type of spin dynamics observed. Below, we present measurements of the ultrafast demagnetization step of every kind of spin-dynamics, excluding type $4$ where no ultrafast demagnetization was detectable, with sub-ps time resolution. The results show that the timescale of ultrafast demagnetization is nearly unaffected.

\begin{figure}[h!]
    \centering
    \includegraphics{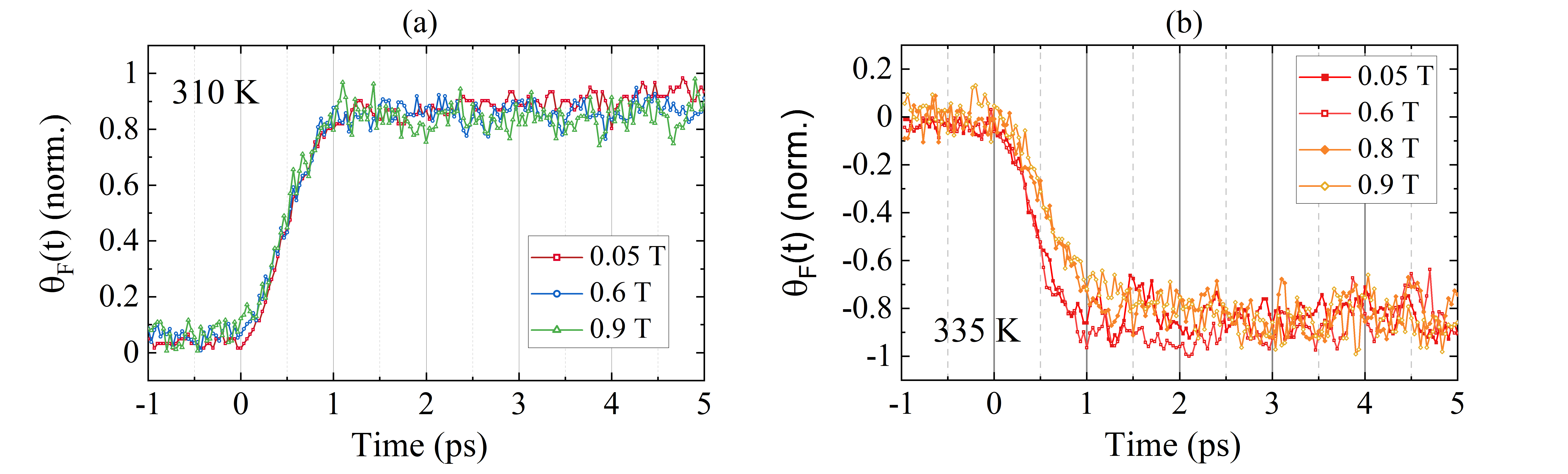}
    \caption{Normalized waveforms of ultrafast demagnetization of any regime of spin dynamics $1$ to $5$ (apart from regime $4$, where ultrafast demagnetization was absent), showing that the timescale is nearly unaffected by the noncollinear arrangement of the spins. (a) was recorded at $T = 310~$K and (b) at $T = 335~$K.}
    \label{fig:UD}
\end{figure}

\bibliography{references}